\documentclass{article}
\usepackage[utf8]{inputenc}
\usepackage[margin=1in]{geometry}
\usepackage[dvipsnames]{xcolor}
\usepackage{amsmath}
\usepackage{graphicx}
\usepackage{amssymb}
\usepackage{enumitem}
\usepackage{mathtools}
\usepackage{authblk}
\usepackage{hyperref}
\usepackage{caption}

\newcommand{\phid}{\hat \phi}
\newcommand{\Md}{\hat M}

\renewcommand{\v}{\overline v}

\renewcommand{\t}{\tau}
\renewcommand{\bold}[1]{\textbf{#1}}
\title{The Glass Transition of Quantum Hard Spheres in High Dimensions}
\author[1]{Michael Winer}
\author[2]{Christopher L. Baldwin}
\author[1]{Richard Barney}
\author[1]{Victor Galitski}
\author[3]{Brian Swingle}
\affil[1]{Joint Quantum Institute, Department of Physics, University of Maryland, College Park, Maryland 20742, USA}
\affil[2]{Department of Physics and Astronomy, Michigan State University, East Lansing, Michigan 48824, USA}
\affil[3]{Department of Physics, Brandeis University, Waltham, Massachusetts 02453, USA}
\date{}

\begin{document}

\maketitle

\begin{abstract}
We study the equilibrium thermodynamics of quantum hard spheres in the infinite-dimensional limit, determining the boundary between liquid and glass phases in the temperature-density plane by means of the Franz-Parisi potential.
We find that as the temperature decreases from high values, the effective radius of the spheres is enhanced by a multiple of the thermal de Broglie wavelength, thus increasing the effective filling fraction and decreasing the critical density for the glass phase. 
Numerical calculations show that the critical density continues to decrease monotonically as the temperature decreases further, suggesting that the system will form a glass at sufficiently low temperatures for any density.
\end{abstract}

\section{Introduction} \label{sec:introduction}

Much of solid-state physics is built around the idea of periodic or almost-periodic crystalline matter. This theory has produced enormous triumphs including the notion of band structure and the Dulong-Petit theory of heat capacity. However, a large amount of solid matter in both the natural and human worlds is instead in a glassy phase --- amorphous matter with no spatial symmetry but with a fixed (or so slowly varying as to effectively be fixed) structure. This phase has a well-studied phenomenology \cite{Berthier,binder2005glassy,Gotze,Lubchenko_2007}, but little can be derived from first principles. The most successful approaches have been mean-field theory \cite{morone2014replica,castellani_2005,Yoshino_2012,altieri2023introduction,mezard1987spin,charbonneau2023spin} and mode-coupling theory \cite{reichman2005mode,szamel2012mode,gotze2009complex,Das,Markland_2011}. 

An important example of the former is the infinite-dimensional glass, which is an exactly solvable model that still displays many of the most important behaviors of glasses, such as arrested dynamics and jamming. The study of the infinite-dimensional glass has seen much progress in recent years, with success in both dynamical~\cite{Liu_2021,Cardenas_1999,Sellitto_2013,maimbourg,Agoritsas_2019} and thermodynamical~\cite{M_zard_1996,Parisi_2010,zamponi2012theory,Berthier2011MicroscopicTO} methods.

Thus far, most of the work done has been for classical glasses.
There has been significant progress studying quantum fluctuations in related problems --- hard-core bosons on the Bethe lattice~\cite{Foini2011Quantum}, the spherical perceptron~\cite{Franz_2019,Artiaco_2021}, ``canyon'' landscapes~\cite{Urbani2023Quantum}, and a larger body of work on quantum spin glasses (see Ref.~\cite{Cugliandolo2022Quantum} and references therein) --- but quantum effects in the infinite-dimensional structural glass itself have not yet been considered.
The purpose of the present paper is to do so, particularly for the prototypical model of hard spheres.

\begin{figure}[h]
\centering
\includegraphics[scale=0.6]{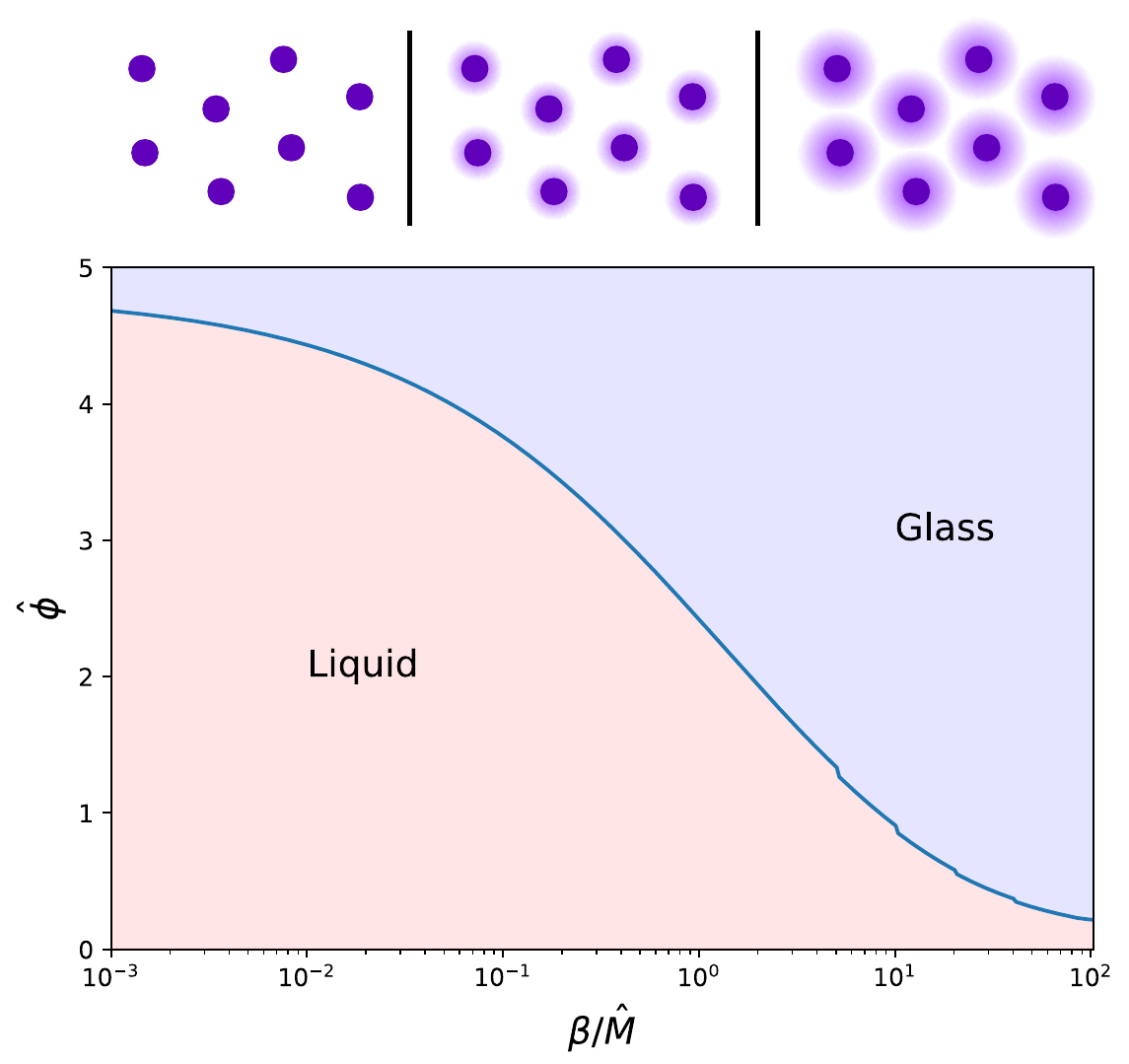}
\caption{Phase diagram of quantum hard spheres in high dimensions as a function of inverse temperature $\beta$ and rescaled filling fraction $\phid$ (see Sec.~\ref{subsec:setup} for the precise definitions of $\phid$ and $\Md$). The relevant qualitative physics is sketched above --- as $\beta$ increases (left to right), quantum fluctuations increase the effective size of the spheres (see also Ref.~\cite{Rozenbaum_2020}) and thus decrease the number of spheres per unit volume needed for glassiness.}
\label{fig:phaseDiagram}
\end{figure}

Of particular relevance for us is Ref.~\cite{resurgence}, which analyzes the mode-coupling theory (MCT) of quantum hard spheres.
MCT is an uncontrolled approximation, but it allows one to directly study the physically relevant three-dimensional system. 
Ref.~\cite{resurgence} finds that weak quantum fluctuations increase the effective radius of the spheres (see the sketch in Fig.~\ref{fig:phaseDiagram}), and thus the critical density for glassiness decreases as one lowers from high temperature.
Interestingly, however, they observe that the liquid phase is re-entrant: as one lowers the temperature, the system will transition from the liquid into the glass phase but then back into the liquid phase.
By contrast, in the infinite-dimensional system considered here, we find that the critical density decreases monotonically as a function of temperature.
We analytically determine the leading-order correction to the classical critical density at high temperature, finding that it is indeed negative (consistent with what is observed in Ref.~\cite{resurgence}), and then numerically determine the phase diagram at much lower temperatures.
The result is shown in Fig.~\ref{fig:phaseDiagram}.
While we cannot reach arbitrarily low temperature in this manner, we do ultimately consider quite small values (namely such that the properly scaled de Broglie wavelength is much larger than the size of the spheres), and still observe a monotonically decreasing critical density.
We do not view these results as incompatible with those of Ref.~\cite{resurgence}, however, since the infinite-dimensional limit and MCT simply entail different approximations.
This is rather an opportunity for further investigation.

\subsection{Definitions and setup} \label{subsec:setup}

We are studying a system of (distinguishable) quantum hard spheres in $d$ dimensions, specifically in the large-$d$ limit.
Some background discussion is given in App.~\ref{app:large_d}.
Our approach is based on the classical work found in, for example, Refs.~\cite{RevModPhys.82.789,parisi2020theory,Parisi_2000,Mari_2011}.

The setup is as follows. We have $N$ spheres labeled by $i \in \{1, \cdots, N\}$, each with radius $\ell/2$ and mass $M$, with momentum operator $\bold P_i$ and position operator $\bold X_i$ (and $\bold X_{ij} \equiv \bold X_i - \bold X_j$). Bold letters denote $d$-dimensional vectors. The system is described by a Hamiltonian
\begin{equation}
    H=\sum_i \frac{\bold P_i^2}{2M}+\sum_{i<j} V(\bold X_{ij}),
    \label{eq:Udef}
\end{equation}
with inverse temperature $\beta$ and particle density $\rho$.
We take the interactions $V$ to be rotationally symmetric but otherwise completely general for as long as possible.
Eventually, however, we will specialize to an infinite hard-sphere potential. In this situation, the closest two spheres can get to each other is $\ell$, with $V(\bold x_{ij}) = \infty$ for $|\bold x_{ij}| < \ell$ and $V(\bold x_{ij}) = 0$ otherwise.

The filling fraction $\phi$ is defined as the product of $\rho$ with the volume of an individual sphere.
It will be convenient to define $\phid \equiv 2^d \phi/d$ and also $\Md \equiv M \ell^2/d^2$. We will find a non-trivial phase diagram when $\phid$ and $\Md$ are held constant as $d$ goes to infinity. This means that in high dimensions, a glass of hard spheres is mostly empty!
The interaction potential $V$ must also scale a certain way with $d$, namely $V(\bold x_{ij}) \equiv \v(d(\frac {|\bold x_{ij}|}\ell-1))$ with $\v(h)$ a $d$-independent function.
Note that a hard-sphere potential corresponds to $\v(h) = \infty$ for $h < 0$ and $\v(h) = 0$ for $h > 0$.

Our analysis will be entirely in terms of a path-integral description of the system.
Thus we will no longer consider operators $\bold P_i$ and $\bold X_i$ but rather trajectories $\bold x_i(\tau)$ with imaginary time $\tau \in [0, \beta)$.
The (Euclidean) Lagrangian is given by
\begin{equation}
\mathcal L = \frac{\Md d^2}{2\ell^2} \sum_i \left| \frac{\textrm{d} \bold x_i(\tau)}{\textrm{d}\tau} \right|^2 + \sum_{i<j} \v \left( d \Big( \frac {|\bold x_{ij}(\tau)|}{\ell} - 1 \Big) \right).
\end{equation}

\subsection{Franz-Parisi potential} \label{subsec:FPP}

We study this system using the Franz-Parisi potential (FPP), a method of extracting dynamical information from purely thermodynamical calculations. Here we give only a brief overview, and refer to Refs.~\cite{parisi2020theory,Franz_1995,Mezard2009} for further details. The FPP is defined as the average free energy of the system when it's constrained to be at a given mean-squared distance $\Delta$ from a reference configuration, with the reference configuration itself drawn from the equilibrium Boltzmann distribution. In equation form, the FPP $\Phi$ is defined as
\begin{equation} \label{eq:FPP_definition}
\beta \Phi \equiv -\int \prod_i \mathcal{D} \bold x_i \frac{e^{-\int \textrm{d}\tau \mathcal{L}[\bold x]}}{Z} \log \int \prod_i \mathcal{D} \bold y_i e^{-\int \textrm{d}\tau \mathcal L[\bold y]} \delta \Big( \frac{1}{N} \sum_i \big| \bold x_i(0) - \bold y_i(0) \big|^2 - \Delta \Big),
\end{equation}
where $Z$ is the partition function and $\prod_i \mathcal{D} \bold x_i$ denotes the path-integral measure.
We will find that $\Phi$ is unbounded from below as $\Delta \rightarrow \infty$, but that it can have a local minimum at finite $\Delta$ for sufficiently large filling fractions.
This is the arrested (i.e., glass) phase --- we refer to the above references for discussion of why this is so.
Thus our aim is specifically to determine whether $\Phi$ defined by Eq.~\eqref{eq:FPP_definition} has a stationary point at finite $\Delta$.

Following Ref.~\cite{parisi2020theory}, we evaluate the FPP using the replica trick and a virial expansion (see Refs.~\cite{castellani_2005,mezard1987spin,Mezard2009,Fischer1991} for details of the former and Refs.~\cite{Hansen2006Theory,kardar_2007,schreiber2011virial,mayer1966statistical} for details of the latter).
The replica trick allows us to write Eq.~\eqref{eq:FPP_definition} as
\begin{equation} \label{eq:FPP_replica_trick}
\begin{aligned}
\beta \Phi &= -\int \prod_i \mathcal{D} \bold x_i \frac{e^{-\int \textrm{d}\tau \mathcal{L}[\bold x]}}{Z} \lim_{m \rightarrow 1} \frac{1}{m-1} \left[ \left( \int \prod_i \mathcal{D} \bold y_i e^{-\int \textrm{d}\tau \mathcal{L}[\bold y]} \delta \Big( \frac{1}{N} \sum_i \big| \bold x_i(0) - \bold y_i(0) \big|^2 - \Delta \Big) \right)^{m-1} - 1 \right] \\
&= -\lim_{m \rightarrow 1} \frac{1}{m-1} \left[ \frac{1}{Z} \int \prod_{ia} \mathcal{D} \bold x_{ia} e^{-\sum_a \int \textrm{d}\tau \mathcal{L}[\bold x_a]} \prod_{a > 1} \delta \Big( \frac{1}{N} \sum_i \big| \bold x_{i1}(0) - \bold x_{ia}(0) \big|^2 - \Delta \Big) - 1 \right],
\end{aligned}
\end{equation}
where to obtain the bottom line, $\bold x_i$ is redefined as $\bold x_{i1}$ and the $m-1$ different factors of $\bold y_i$ are redefined as $\bold x_{ia}$ for $a \in \{2, \cdots, m\}$ (thus in total $a$ ranges from 1 to $m$).
We thus need only to calculate the ``replicated'' partition function
\begin{equation} \label{eq:replicated_partition_function_definition}
Z_{\textrm{rep}} \equiv \int \prod_{ia} \mathcal{D} \bold x_{ia} e^{-\sum_a \int \textrm{d}\tau \mathcal{L}[\bold x_a]} \prod_{a > 1} \delta \Big( \frac{1}{N} \sum_i \big| \bold x_{i1}(0) - \bold x_{ia}(0) \big|^2 - \Delta \Big).
\end{equation}
Eq.~\eqref{eq:FPP_replica_trick} then states that $Z_{\textrm{rep}} \sim Z [1 - (m-1) \beta \Phi]$ for $m$ close to 1, i.e., $-\beta \Phi$ is the first-order coefficient in the expansion of $\log{Z_{\textrm{rep}}}$ around $m = 1$.

We calculate $\log{Z_{\textrm{rep}}}$ via a virial expansion~\cite{Hansen2006Theory,kardar_2007,schreiber2011virial,mayer1966statistical}.
This expansion is in terms of the single-particle density $\rho(\bold x)$, and thus the resulting free energy includes $\rho(\bold x)$ as a variational function to be minimized over.
Since we are considering a replicated quantum system, $\rho(\bold x)$ is still a function of the coordinates of all replicas at all imaginary times, i.e., the argument $\bold x$ in $\rho(\bold x)$ has replica and imaginary-time indices --- $\bold x \equiv \{ \bold x_a(\tau) \}_{a, \tau}$.
Yet importantly, it no longer has a particle index $i$.
Furthermore, taking the large-$d$ limit allows us to truncate the expansion at second order~\cite{parisi2020theory}.
The replicated free energy then amounts to (up to an overall constant concerning normalization of the path integral)
\begin{equation} \label{eq:replicated_free_energy_virial_expansion}
\begin{aligned}
\beta f \equiv -\frac{\log{Z_{\textrm{rep}}}}{Nd} &= \min_{\rho} \left\{ \frac{1}{Nd} \int \prod_a \mathcal{D} \bold x_a \rho(\bold x) \left[ \log{\rho(\bold x)} + \frac{M}{2} \sum_a \int_0^{\beta} \textrm{d}\tau \left| \frac{\textrm{d} \bold x_a(\tau)}{\textrm{d}\tau} \right|^2 \right] \right. \\
&\qquad \left. - \frac{1}{2Nd} \int \prod_a \mathcal{D} \bold x_a \mathcal{D} \bold y_a \rho(\bold x) \rho(\bold y) \left( \exp \left[ -\sum_a \int_0^{\beta} \textrm{d}\tau V \big( \bold x_a(\tau) - \bold y_a(\tau) \big) \right] - 1 \right) \right\}.
\end{aligned}
\end{equation}
Keep in mind that we still have the constraint $N^{-1} \sum_i |\bold x_{i1}(0) - \bold x_{ia}(0)|^2 = \Delta$.
This is easy to account for in Eq.~\eqref{eq:replicated_free_energy_virial_expansion} --- only consider densities $\rho(\bold x)$ such that $|\bold x_1(0) - \bold x_a(0)|^2 = \Delta$ in the minimization.

An additional simplification that occurs at large $d$ is that all terms in Eq.~\eqref{eq:replicated_free_energy_virial_expansion} depend only on the ``second moment'' of $\rho(\bold x)$ (interpreting $\rho(\bold x)/N$ as a probability distribution on $\bold x$)~\cite{parisi2020theory}.
Thus rather than minimize over all functions $\rho(\bold x)$, we need only minimize over its covariance matrix.

Even this can be further simplified by making use of the symmetries in the system.
Here it is helpful to think of the $m$ replicas of each sphere as forming a ``molecule''.
Let us first consider the corresponding classical system --- each molecule is characterized by $md$ numbers (one for each component of each $\bold x_a$), and in principle $\rho(\bold x)$ could depend on all of them.
Yet since the system is translation- and rotation-invariant, $\rho(\bold x)$ in fact depends only on the $m(m+1)/2$ numbers in the matrix $(\bold x_a - \overline{\bold x}) \cdot (\bold x_b - \overline{\bold x})$, where $\overline{\bold x} \equiv m^{-1} \sum_c \bold x_c$ is the ``center of mass'' of the molecule.
Furthermore, as noted above, the replicated free energy depends only on the expectation value of this quantity, denoted $\alpha_{ab}$:
\begin{equation} \label{eq:classical_alpha_definition}
\alpha_{ab} \equiv \frac{d}{\ell^2} \big< (\bold x_a - \overline{\bold x}) \cdot (\bold x_b - \overline{\bold x}) \big>,
\end{equation}
where the brackets denote an average with respect to $\rho(\bold x)$.
The explicit factors of $d$ and $\ell$ are chosen so that $\alpha_{ab}$ is $O(1)$ in the large-$d$ limit.
Note that even the entries of $\alpha$ are not entirely independent, since by definition the all-1 vector has eigenvalue 0: $\sum_b (\bold x_b - \overline{\bold x}) = 0$ and thus $\sum_b \alpha_{ab} = 0$.
Also note that the constraint amounts to requiring $\alpha_{11} + \alpha_{aa} - 2 \alpha_{a1} = d \Delta / \ell^2$ for all $a > 1$.

Recall that once we calculate the replicated free energy $f$ and then the FPP $\Phi$, our goal is to identify points where $\textrm{d}\Phi / \textrm{d}\Delta = 0$.
At such points, if they exist, we have a further permutation symmetry between replicas (see Ref.~\cite{parisi2020theory}).
Thus there are only two distinct elements of $\alpha$ --- the diagonal and off-diagonal entries --- and the fact that $\sum_b \alpha_{ab} = 0$ fixes one in terms of the other.
We can therefore parametrize $\alpha$ by a single number $A$ (with this specific parametrization chosen so as to match the expressions in Ref.~\cite{parisi2020theory}):
\begin{equation} \label{eq:classical_alpha_form}
\alpha = \frac{1}{2m} \begin{pmatrix} (m-1)A & -A & \cdots & -A \\ -A & (m-1)A & \cdots & -A \\ \vdots & \vdots & \ddots & \vdots \\ -A & -A & \cdots & (m-1)A \end{pmatrix}.
\end{equation}

When we promote from classical to quantum thermodynamics, the $md$ coordinates of each molecule become $md$ imaginary-time trajectories.
The matrix $\alpha$ then acquires time dependence as well:
\begin{equation} \label{eq:quantum_alpha_definition}
\alpha_{ab}(\tau_1, \tau_2) \equiv \frac{d}{\ell^2} \big< \big( \bold x_a(\tau_1) - \overline{\bold x} \big) \cdot \big( \bold x_b(\tau_2) - \overline{\bold x} \big) \big>,
\end{equation}
where, analogous to above, $\overline{\bold x} \equiv (m\beta)^{-1} \int \textrm{d}\tau \sum_c \bold x_c(\tau)$.
However, each replica has a separate time-translation symmetry, meaning that $\langle \bold x_a(\tau_1) \cdot \bold x_a(\tau_2) \rangle$ depends only on $\tau_1 - \tau_2 \equiv \tau_{12}$ and $\langle \bold x_a(\tau_1) \cdot \bold x_b(\tau_2) \rangle$ is independent of both $\tau_1$ and $\tau_2$.
Accounting for replica symmetry as well, this means that $\alpha$ takes the form
\begin{equation} \label{eq:quantum_alpha_form}
\alpha(\tau_1, \tau_2) = \frac{1}{2m} \begin{pmatrix} (m-1)A + mG(\tau_{12}) & -A & \cdots & -A \\ -A & (m-1)A + mG(\tau_{12}) & \cdots & -A \\ \vdots & \vdots & \ddots & \vdots \\ -A & -A & \cdots & (m-1)A + mG(\tau_{12}) \end{pmatrix}.
\end{equation}
As before, $\int \textrm{d}\tau_2 \sum_b \alpha_{ab}(\tau_1, \tau_2) = 0$, which translates to $\int \textrm{d}\tau G(\tau) = 0$.
In other words, we are fixing the zero-frequency component of $G_{\omega} \equiv \int \textrm{d}\tau e^{i \omega \tau} G(\tau)$ to be identically zero.

\subsection{Plan for the paper}

To summarize, our strategy for determining the phase diagram of the quantum hard-sphere liquid is as follows:
\begin{itemize}
\item Express the replicated free energy given by Eq.~\eqref{eq:replicated_free_energy_virial_expansion} in terms of the dynamical correlation function $G(\tau)$ (equivalently $G_{\omega}$) and the inter-replica overlap $A$.
\item Obtain saddle-point equations for $G_{\omega}$ and $A$ in the limit $m \rightarrow 1$.
\item We will find that the equations for $G_{\omega}$ do not depend on $A$ (and in fact require only the zeroth-order term of $f$). Thus solve for $G_{\omega}$ first.
\item Insert the solution for $G_{\omega}$ into the equation for $A$ (which does require the first-order term of $f$, i.e., the FPP $\Phi$). We will find that $A = \infty$ is always a solution. If this is the only solution, then the system is in the liquid phase. If there is also a finite-$A$ solution, on the other hand, then the system is in the glass phase.
\end{itemize}

One might notice that in this procedure, we never actually calculate the FPP as a function of the mean-squared distance $\Delta$.
This is because, as stated above, determining the phase diagram only requires knowledge of the stationary points of the FPP.
We can perform the extremization over $\Delta$ at the same time as we extremize over the other quantities needed to determine the free energy, and within the replica-symmetric ansatz described above, this amounts to extremizing over the parameter $A$.

In the remainder of the paper, we first evaluate the replicated free energy and determine its saddle-point equations in Sec.~\ref{sec:action}.
We analytically solve the equations at high temperature in Sec.~\ref{sec:high_temperatures}, finding that quantum fluctuations ``renormalize'' the radius of the spheres to be slightly larger (and thus the glass phase appears at a slightly lower density).
We then numerically solve the equations at lower temperatures in Sec.~\ref{sec:numerics}.
We do not observe any sign of a re-entrant liquid phase --- the critical density instead decreases monotonically as the temperature lowers.
Lastly, we discuss future directions in Sec.~\ref{sec:conclusion}.

\section{The free energy and its saddle points} \label{sec:action}

Here we evaluate the replicated free energy given in Eq.~\eqref{eq:replicated_free_energy_virial_expansion}.
It is helpful to do so in stages, first determining the classical free energy (which will be entirely a review of known results) and then incorporating quantum fluctuations.

\subsection{Classical free energy} \label{subsec:classicalAction}

Even for the classical free energy, we proceed in stages: first calculate the unreplicated non-interacting free energy, then the unreplicated interacting free energy, and lastly the replicated interacting free energy.

In the absence of interactions and without any replicas, the free energy is trivial to evaluate --- in any number of dimensions, the partition function is $Z = V^N / N!$, where $V$ denotes the volume.
This gives a free energy $f_{\textrm{position}}$:
\begin{equation} \label{eq:freeEnGas}
\beta f_{\textrm{position}} = -\frac{\log{Z}}{Nd} \sim \frac{\log \rho}{d} \sim \frac{1}{2} \log \frac{d}{2\pi e \ell^2},
\end{equation}
where we've neglected terms that vanish at large $d$ (recall that $\phid \equiv 2^d \phi/d$ is kept $O(1)$ at large $d$ and $\phi \sim (\pi e \ell^2 / 2d)^{d/2} \rho$ --- see App.~\ref{app:large_d}).
Note that $\beta f_{\textrm{position}}$ is equivalently the contribution of the first term in Eq.~\eqref{eq:replicated_free_energy_virial_expansion} --- for a translation-invariant classical system without replicas, $\rho(\bold x)$ is simply a constant $\rho$ and so $\int \textrm{d} \bold x \rho(\bold x) \log{\rho(\bold x)} = N \log{\rho}$.
Also note that this contribution depends logarithmically on $d$.
Usually the free energy is proportional to the number of degrees of freedom, but in the limit we are taking there is this additional term going as $\log d$.

Now include interactions (still without replicas).
The correction coming from the virial expansion is
\begin{equation} \label{eq:simpleVirial}
\beta f_{\textrm {virial}} = -\frac{\rho}{2d} \int \textrm{d} \bold x \left( e^{-\beta V(\bold x)}-1\right) = -\frac{\phid}{2} \int  \textrm{d}h e^h \left( e^{-\beta \v(h)} - 1 \right), 
\end{equation}
where the final expression follows from writing $V(\bold x) = \bar v (d(|\bold x|/\ell-1))$, substituting $h=d(|\bold x|/\ell-1)$, and using that $|\bold x|^{d-1} \sim \ell^{d-1} e^{h}$ at large $d$.
For the case of hard spheres ($\v (h) = \infty$ for $h < 0$ and $\v (h) = 0$ for $h > 0$), the integral in Eq.~\eqref{eq:simpleVirial} evaluates to $-1$ and thus
\begin{equation}
\beta f_{\textrm {virial}} = \frac{\phid}{2}.
\end{equation}

Now introduce replicas.
As mentioned in Sec.~\ref{sec:introduction}, it is convenient to think of each set of $m$ replicas as forming a single ``molecule''.
In calculating the entropic contribution to the free energy (first term of Eq.~\eqref{eq:replicated_free_energy_virial_expansion}), we still obtain a term $\beta f_{\textrm{position}}$ coming from the overall density of molecules.
Yet now the internal entropy of each molecule is non-trivial.
As noted above, the other terms of Eq.~\eqref{eq:replicated_free_energy_virial_expansion} depend only on the covariance matrix $\ell^2 \alpha/d^2$ of the coordinates $\bold x_a - \overline{\bold x}$.
Thus in minimizing the free energy, we should take $\rho(\bold x)$ to be the distribution with maximum entropy subject to fixed covariance, namely Gaussian.
Using the standard result for the entropy of a Gaussian, we obtain a ``molecular shape'' contribution to the free energy given by
\begin{equation} \label{eq:classical_molecular_shape_contribution_v1}
\beta f_{\textrm{molecular}} = -\frac{1}{2} \log{\textrm{Det}' \frac{2\pi e \ell^2 \alpha}{d^2}}, 
\end{equation}
where $\textrm{Det}'$ indicates that the zero eigenvalue of $\alpha$ is omitted.
Given the form in Eq.~\eqref{eq:classical_alpha_form}, this evaluates to
\begin{equation} \label{eq:classical_molecular_shape_contribution_v2}
\beta f_{\textrm{molecular}} = -\frac{m-1}{2} \log{\frac{\pi e \ell^2 A}{d^2}}.
\end{equation}

Calculating the interaction term in the replicated setting requires more thought.
Referring to Eq.~\eqref{eq:replicated_free_energy_virial_expansion}, we see that we need to average $\exp [-\beta \sum_a V(\bold x_a - \bold y_a)]$ over the coordinates of two molecules $\bold x$ and $\bold y$.
The relative coordinates $\bold x_a - \overline{\bold x}$ and $\bold y_a - \overline{\bold y}$ are Gaussian-distributed with covariance matrix $\ell^2 \alpha/d^2$, while $\overline{\bold x}$ and $\overline{\bold y}$ are uniformly distributed by translation invariance.
Thus define $\bold u_a \equiv \bold x_a - \overline{\bold x}$ and $\bold v_a \equiv \bold y_a - \overline{\bold y}$, as well as $\bold r \equiv \overline{\bold x} - \overline{\bold y}$.
The interaction term then amounts to
\begin{equation} \label{eq:classical_replicated_virial_formal_form}
\beta f_{\textrm{virial}} = -\frac{\rho}{2d} \int \textrm{d} \bold r \left( \big< e^{-\beta \sum_a V(\bold u_a - \bold v_a + \bold r)} \big> - 1 \right),
\end{equation}
where the brackets denote an average over the aforementioned Gaussians $\bold u$ and $\bold v$.

For the radially symmetric interactions considered here, so that $V(\bold u_a - \bold v_a + \bold r)$ depends only on $|\bold u_a - \bold v_a + \bold r|$, we can substantially simplify further.
Writing the separation as
\begin{equation} \label{eq:distance_expansion}
|\bold u_a - \bold v_a + \bold r| = \sqrt{|\bold r|^2 + 2 \bold r \cdot (\bold u_a - \bold v_a) + |\bold u_a - \bold v_a|^2} \sim |\bold r| + \frac{\bold r \cdot (\bold u_a - \bold v_a)}{|\bold r|} + \frac{|\bold u_a - \bold v_a|^2}{2 |\bold r|},
\end{equation}
the term $\bold r \cdot (\bold u_a - \bold v_a) / |\bold r|$ is a Gaussian with covariance $2 \ell^2 \alpha_{ab}/d^2$, and the term $|\bold u_a - \bold v_a|^2 / 2 |\bold r|$ is deterministic and equal to $\ell^2 \alpha_{aa}/d |\bold r|$ at large $d$.
Thus
\begin{equation} \label{eq:classical_replicated_virial_inner_average}
\big< e^{-\beta \sum_a V(\bold u_a - \bold v_a + \bold r)} \big> = \big( \textrm{Det} 4\pi \alpha \big)^{-\frac{1}{2}} \int \prod_a \textrm{d}z_a e^{-\frac{1}{4} \sum_{ab} \alpha_{ab}^{-1} z_a z_b} \prod_a e^{-\beta \v (h + \alpha_{aa} + z_a)},
\end{equation}
where we have again substituted $h=d(|\bold r|/\ell-1)$.
Using the identity~\cite{parisi2020theory}
\begin{equation} \label{eq:Gaussian_convolution_identity}
\big( \textrm{Det} 4\pi \alpha \big)^{-\frac{1}{2}} \int \prod_a \textrm{d}z_a e^{-\frac{1}{4} \sum_{ab} \alpha_{ab}^{-1} z_a z_b} \prod_a e^{-\beta \v (h + \alpha_{aa} + z_a)} = e^{\sum_{ab} \alpha_{ab} \frac{\partial^2}{\partial h_a \partial h_b}} \prod_a e^{-\beta \v (h_a)} \bigg|_{h_a = h + \alpha_{aa}},
\end{equation}
we can thus write the interaction term as
\begin{equation} \label{eq:classical_replicated_virial_term_v1}
\beta f_{\textrm{virial}} = -\frac{\phid}{2} \int \textrm{d}h e^h \left[ e^{\sum_{ab} \alpha_{ab} \frac{\partial^2}{\partial h_a \partial h_b}} \prod_a e^{-\beta \v (h_a)} \bigg|_{h_a = h + \alpha_{aa}} - 1 \right].
\end{equation}
Given the form of $\alpha$ in Eq.~\eqref{eq:classical_alpha_form}, $\beta f_{\textrm{virial}}$ further simplifies to
\begin{equation} \label{eq:classical_replicated_virial_term_v2}
\begin{aligned}
\beta f_{\textrm{virial}} &= -\frac{\phid}{2} \int \textrm{d}h e^h \left[ e^{-\frac{A}{2m} \frac{\textrm{d}^2}{\textrm{d}h^2}} \prod_a e^{\frac{A}{2} \frac{\partial^2}{\partial h_a^2}} e^{-\beta \v (h_a)} \bigg|_{h_a = h + \frac{(m-1)A}{2m}} - 1 \right] \\
&= -\frac{\phid}{2} \int \textrm{d}h \left[ \left( e^{-\frac{A}{2m} \frac{\textrm{d}^2}{\textrm{d}h^2}} e^h \right) \prod_a e^{\frac{A}{2} \frac{\partial^2}{\partial h_a^2}} e^{-\beta \v (h_a)} \bigg|_{h_a = h + \frac{(m-1)A}{2m}} - e^h \right] \\
&= -\frac{\phid}{2} \int \textrm{d}h e^h \left[ \left( e^{\frac{A}{2} \frac{\textrm{d}^2}{\textrm{d}h^2}} e^{-\beta \v (h + \frac{A}{2})} \right)^m - 1 \right].
\end{aligned}
\end{equation}

All in all, the replicated free energy of the classical liquid is given by the sum of $f_{\textrm{position}}$ (Eq.~\eqref{eq:freeEnGas}), $f_{\textrm{molecular}}$ (Eq.~\eqref{eq:classical_molecular_shape_contribution_v2}), and $f_{\textrm{virial}}$ (Eq.~\eqref{eq:classical_replicated_virial_term_v2}):
\begin{equation} \label{eq:classical_replicated_free_energy_result}
\beta f_{\textrm{classical}} = \frac{1}{2} \log \frac{d}{2\pi e \ell^2} - \frac{m-1}{2} \log{\frac{\pi e \ell^2 A}{d^2}} - \frac{\phid}{2} \int \textrm{d}h e^h \left[ \left( e^{\frac{A}{2} \frac{\textrm{d}^2}{\textrm{d}h^2}} e^{-\beta \v (h + \frac{A}{2})} \right)^m - 1 \right].
\end{equation}

\subsection{Quantum free energy} \label{subsec:quantumAction}

Moving on to the quantum system, it is important to balance two interpretations.
In one, we have $m$ replicas, each completing periodic motion on the thermal circle with period $\beta$ under the influence of kinetic and potential energies.
Following this approach, one could recalculate all of the above terms while keeping track of additional imaginary-time indices.
The only wholly new contribution to the free energy is the kinetic term (second term of the first line in Eq.~\eqref{eq:replicated_free_energy_virial_expansion}), which evalutes to
\begin{equation} \label{eq:quantum_kinetic_term}
\beta f_{\textrm{kinetic}} = \frac{M \ell^2}{2d^2} \sum_a \int_0^{\beta} \textrm{d}\tau \frac{\partial^2}{\partial \tau_1 \partial \tau_2} \alpha_{aa}(\tau_1, \tau_2) \Big|_{\t_1,\t_2=\t} = -\frac{\Md m \beta}{4} G''(0).
\end{equation}

The other interpretation, however, is to divide imaginary time into $\beta / \Delta \tau$ slices and treat the system as having $m \beta / \Delta \tau$ replicas, with the kinetic term simply as an interaction between them.
With this interpretation, it is natural that $\alpha_{ab}$ should be promoted to $\alpha_{ab}(\tau_1, \tau_2)$ but otherwise be treated as before.
We can then easily write down how the classical free energy should be modified (and we have confirmed that a more mechanical derivation gives the same result).

The term $f_{\textrm{position}}$ is the same as in the classical case, and the term $f_{\textrm{molecular}}$ simply gains a contribution coming from non-zero frequencies (note that the zero-frequency contribution is unmodified since we have fixed $G_0 = 0$):
\begin{equation} \label{eq:quantum_molecular_shape_contribution}
\beta f_{\textrm{molecular}} = -\frac{m-1}{2} \log{\frac{\pi e \ell^2 A}{d^2}} - \frac{m}{2} \sum_{\omega \neq 0} \log{G_{\omega}}.
\end{equation}
The interaction term becomes
\begin{equation} \label{eq:quantum_interaction_term}
\begin{aligned}
\beta f_{\textrm{virial}} &= -\frac{\phid}{2} \int \textrm{d}h e^h \left[ e^{\sum_{ab} \int \textrm{d}\tau_1 \textrm{d}\tau_2 \alpha_{ab}(\tau_{12}) \frac{\partial^2}{\partial h_a(\tau_1) \partial h_b(\tau_2)}} \prod_a e^{-\int \textrm{d}\tau \v (h_a(\tau))} \bigg|_{h_a(\tau) = h + \alpha_{aa}(0)} - 1 \right] \\
&= -\frac{\phid}{2} \int \textrm{d}h e^h \left[ \left( e^{\int \textrm{d}\tau_1 \textrm{d}\tau_2 \frac{A + G(\tau_{12})}{2} \frac{\partial^2}{\partial h(\tau_1) \partial h(\tau_2)}} e^{-\int \textrm{d}\tau \v (h(\tau))} \bigg|_{h(\tau) = h + \frac{A + G(0)}{2}} \right)^m - 1 \right].
\end{aligned}
\end{equation}
Thus the replicated free energy of the quantum liquid, i.e., Eq.~\eqref{eq:replicated_free_energy_virial_expansion}, is
\begin{equation} \label{eq:quantum_replicated_free_energy}
\begin{aligned}
\beta f &= \frac{1}{2} \log \frac{d}{2\pi e \ell^2} - \frac{m-1}{2} \log{\frac{\pi e \ell^2 A}{d^2}} + \frac{m}{2} \sum_{\omega \neq 0} \left( \frac{\Md}{2} \omega^2 G_{\omega} - \log{G_{\omega}} \right) \\
&\qquad - \frac{\phid}{2} \int \textrm{d}h e^h \left[ \left( e^{\int \textrm{d}\tau_1 \textrm{d}\tau_2 \frac{A + G(\tau_{12})}{2} \frac{\partial^2}{\partial h(\tau_1) \partial h(\tau_2)}} e^{-\int \textrm{d}\tau \v (h(\tau))} \bigg|_{h(\tau) = h + \frac{A + G(0)}{2}} \right)^m - 1 \right].
\end{aligned}
\end{equation}

We can rewrite Eq.~\eqref{eq:quantum_replicated_free_energy} using some convenient notation.
Let $\langle \, \cdot \, \rangle_{r,GA}$ denote the average over a periodic Gaussian process $r(\tau)$ with covariance matrix $\langle r(\tau_1) r(\tau_2) \rangle_{r,GA} = A + G(\tau_{12})$.
By Eq.~\eqref{eq:Gaussian_convolution_identity}, the term in parentheses in the second line of Eq.~\eqref{eq:quantum_replicated_free_energy} is $\langle \exp [-\int \textrm{d}\tau \v (h + r(\tau))] \rangle_{r,GA}$ evaluated at $h + (A + G(0))/2$.
Denote this quantity by $p_{GA}(h)$:
\begin{equation} \label{eq:Brownian_probability_definition}
p_{GA}(h) \equiv \Big< e^{-\int \textrm{d}\tau \v (h + r(\tau))} \Big>_{r,GA},
\end{equation}
so that we can write
\begin{equation} \label{eq:quantum_replicated_free_energy_compact}
\beta f = \frac{1}{2} \log \frac{d}{2\pi e \ell^2} - \frac{m-1}{2} \log{\frac{\pi e \ell^2 A}{d^2}} + \frac{m}{2} \sum_{\omega \neq 0} \left( \frac{\Md}{2} \omega^2 G_{\omega} - \log{G_{\omega}} \right) - \frac{\phid}{2} \int \textrm{d}h e^{h - \frac{A + G(0)}{2}} \Big[ p_{GA}(h)^m - 1 \Big]. \; \;
\end{equation}
An identity that will often be useful is that
\begin{equation} \label{eq:Brownian_probability_identity}
p_{GA}(h) = e^{\frac{A}{2} \frac{\textrm{d}^2}{\textrm{d}h^2}} p_{G0}(h) = \int \frac{\textrm{d}z}{\sqrt{2\pi A}} e^{-\frac{z^2}{2A}} p_{G0}(h+z).
\end{equation}

\subsection{Saddle points} \label{subsec:saddle_points}

Our goal is to extremize Eq.~\eqref{eq:quantum_replicated_free_energy_compact} in the limit $m \rightarrow 1$.
As described in Sec.~\ref{subsec:FPP}, the FPP $\Phi$ in particular is the first-order coefficient of the free energy: $f \sim f_0 + (m-1) \Phi$.
However, care must be taken in performing the extremization.
If the zeroth-order term $f_0$ has non-trivial dependence on a parameter, then extremizing $f_0$ alone determines that parameter to leading order.
Yet if $f_0$ is independent of it, then extremizing $\Phi$ instead determines the parameter to leading order.
In our case, we have that (neglecting overall constants)
\begin{align}
\beta f_0 &= \frac{1}{2} \sum_{\omega \neq 0} \left( \frac{\Md}{2} \omega^2 G_{\omega} - \log{G_{\omega}} \right) - \frac{\phid}{2} \int \textrm{d}h e^{h - \frac{G(0)}{2}} \Big[ p_{G0}(h) - 1 \Big], \label{eq:zeroth_order_quantum_free_energy} \\
\beta \Phi &= - \frac{1}{2} \log{A} + \frac{1}{2} \sum_{\omega \neq 0} \left( \frac{\Md}{2} \omega^2 G_{\omega} - \log{G_{\omega}} \right) - \frac{\phid}{2} \int \textrm{d}h e^{h - \frac{A + G(0)}{2}} p_{GA}(h) \log{p_{GA}(h)}. \label{eq:first_order_quantum_free_energy}
\end{align}
Note that $f_0$ is independent of $A$.
Thus we should extremize $f_0$ with respect to $G_{\omega}$ and extremize $\Phi$ with respect to $A$.

Setting $\partial f_0 / \partial G_{\omega} = 0$ gives the following saddle-point equations:
\begin{equation} \label{eq:G_saddle_point_equations}
\begin{aligned}
0 &= \frac{\Md \omega^2}{4} - \frac{1}{2 G_{\omega}} + \frac{\phid}{4 \beta} \int \textrm{d}h e^{h - \frac{G(0)}{2}} \Big[ p_{G0}(h) - 1 \Big] \\
&\qquad + \frac{\phid}{4 G_{\omega}} \int \textrm{d}h e^{h - \frac{G(0)}{2}} p_{G0}(h) - \frac{\phid}{4 G_{\omega}^2} \int \textrm{d}h e^{h - \frac{G(0)}{2}} \Big< r_{\omega} r_{-\omega} e^{-\int \textrm{d}\tau \v (h + r(\tau))} \Big>_{r,G0}.
\end{aligned}
\end{equation}
While intimidating, Eq.~\eqref{eq:G_saddle_point_equations} can be solved numerically by iteration as described in Sec.~\ref{sec:numerics}.

Once Eq.~\eqref{eq:G_saddle_point_equations} is solved, setting $\partial \Phi / \partial A = 0$ gives the following saddle-point equation for $A$:
\begin{equation} \label{eq:A_saddle_point_equation}
0 = -\frac{1}{2A} + \frac{\phid}{4} \int \textrm{d}h e^{h - \frac{A + G(0)}{2}} p_{GA}(h) \log{p_{GA}(h)} - \frac{\phid}{4} \int \textrm{d}h e^{h - \frac{A + G(0)}{2}} p_{GA}''(h) \Big[ \log{p_{GA}(h)} + 1 \Big].
\end{equation}
We can integrate by parts to write this more compactly as
\begin{equation} \label{eq:A_saddle_point_equation_compact}
\frac{1}{A} = \frac{\phid}{2} \int \textrm{d}h e^{h - \frac{A + G(0)}{2}} \frac{p_{GA}'(h)^2}{p_{GA}(h)}.
\end{equation}
Keeping in mind that $G_{\omega}$ has already been determined, Eq.~\eqref{eq:A_saddle_point_equation_compact} is a single equation for $A$ that can simply be plotted to determine graphically whether solutions exist.

\section{High Temperatures} \label{sec:high_temperatures}

Here we at last specialize to hard spheres, corresponding to $\v(h) = \infty$ for $h < 0$ and $\v(h) = 0$ for $h > 0$.
This gives $p_{GA}(h)$ as expressed in Eq.~\eqref{eq:Brownian_probability_definition} a particularly simple interpretation: it is the probability that the Gaussian process $r(\tau)$ never crosses below $-h$, which is equivalently the probability that the \textit{minimum} of the process is greater than $-h$.
By symmetry, $p_{GA}(h)$ is also the probability that the maximum is less than $h$, and therefore $p_{GA}'(h)$ is the probability density function for $\max_{\tau} r(\tau)$.

\begin{figure}
\centering
\includegraphics[width=0.8\textwidth]{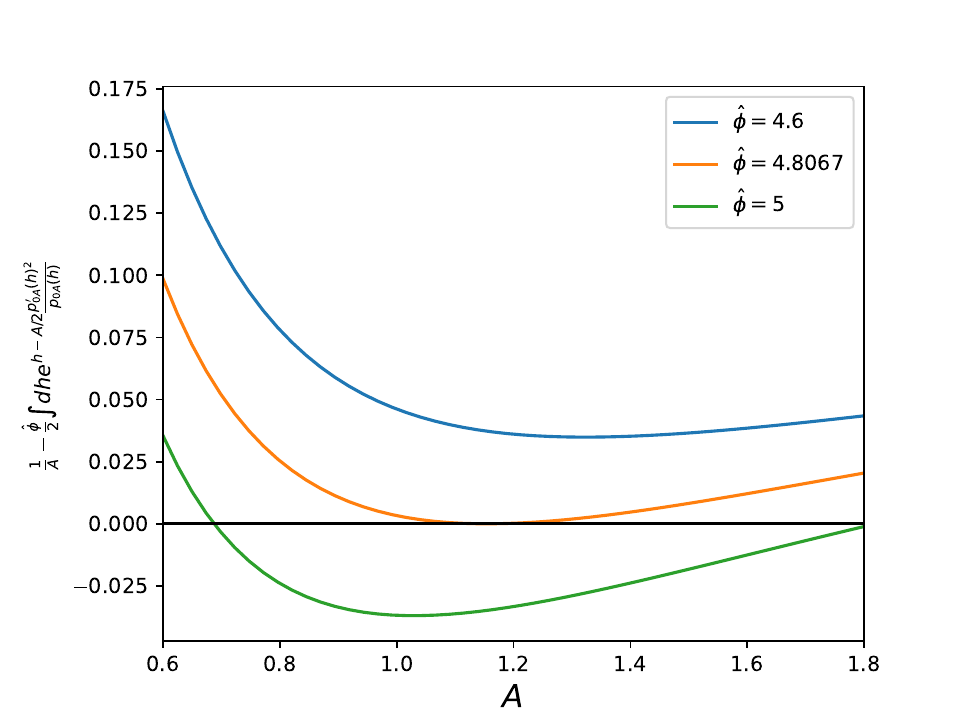}
\caption{Eq.~\eqref{eq:A_saddle_point_equation_compact} (left-hand minus right-hand side) as a function of $A$ for different values of $\phid$, specifically in the infinite-temperature (i.e., classical) limit. At the critical value $\phid_c \approx 4.8067$, the curve acquires a zero near $A \approx 1.15$, indicating that the system is in a glass phase.}
\label{fig:classicalGraph}
\end{figure}

As a warm-up, consider the saddle-point equations at infinite temperature, which corresponds to the classical limit.
Then $G_{\omega} = 0$, there is no $\tau$-dependence to account for, and $r$ is simply a Gaussian with variance $A$:
\begin{equation} \label{eq:Brownian_probability_classical_result}
p_{0A}(h) = \int_{-h}^{\infty} \frac{\textrm{d}r}{\sqrt{2\pi A}} e^{-\frac{r^2}{2A}}.
\end{equation}
We insert this expression into Eq.~\eqref{eq:A_saddle_point_equation_compact}.
A graphical inspection as in Fig.~\ref{fig:classicalGraph} demonstrates that solutions exist for $\phid > \phid_c \approx 4.8067$.
This is the glass phase of the system.

At large but finite temperatures, $r(\tau)$ is no longer constant.
Since fluctuations in time can only increase the probability that $\max_{\tau} r(\tau)$ exceeds some value, $p_{GA}(h) < p_{0A}(h)$.
We will show that $p_{GA}(h)$ can be well-approximated by $p_{0A}(h - \delta R)$ at high temperatures for some $\delta R > 0$.
Our strategy is to determine $p_{G0}(h)$ and then convolve with a Gaussian of variance $A$ to obtain $p_{GA}(h)$ (see Eq.~\eqref{eq:Brownian_probability_identity}).

\begin{figure}
\centering
\includegraphics[scale=0.7]{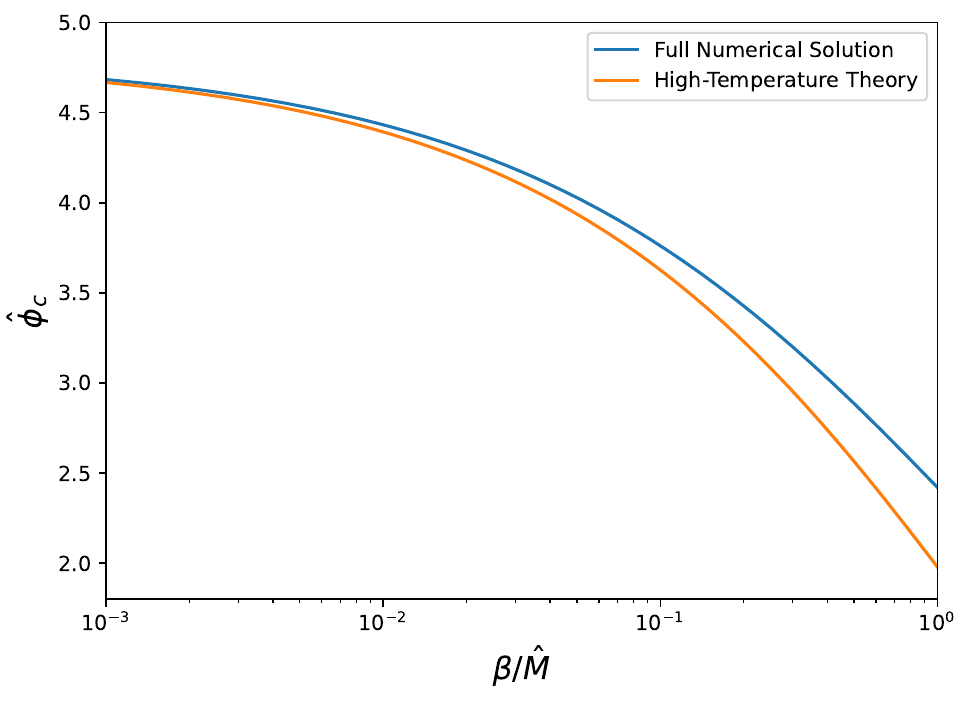}
\caption{Comparison of $\phid_c(\beta)$ estimated analytically at high temperatures (orange curve --- see Sec.~\ref{sec:high_temperatures}) to that obtained numerically (blue curve --- see Sec.~\ref{sec:numerics}).}
\label{fig:smallBetaGraph}
\end{figure}

The first thing to note is that at high temperatures,
\begin{equation} \label{eq:G_high_temperature_limit}
G_{\omega} \sim \frac{2}{\Md \omega^2},
\end{equation}
keeping in mind that $G_0 = 0$.
First and foremost, this is confirmed by a full numerical solution of the saddle-point equations as discussed in Sec.~\ref{sec:numerics} --- Fig.~\ref{fig:smallBetaGraph} shows that the critical density $\phid_c(\beta)$ obtained by assuming Eq.~\eqref{eq:G_high_temperature_limit} matches the numerical result as $\beta \rightarrow 0$.
It can also be justified by returning to Eq.~\eqref{eq:zeroth_order_quantum_free_energy} for $f_0$ --- whereas each of the first two terms has a derivative going as $O(\beta^{-2})$, the last term is exponential in $G$ (recall that $p_{G0}(h)$ can be written as an exponential of second derivatives) and so its derivative is at most only $O(\beta^{-1})$ (the factor of $\beta^{-1}$ because $G(\tau) = \beta^{-1} \sum_{\omega} e^{-i \omega \tau} G_{\omega}$).

In any case, note that Eq. \eqref{eq:G_high_temperature_limit} is the covariance for Brownian motion, i.e., the action governing the distribution of $r(\tau)$ is simply $\frac{\Md}{4} \int \textrm{d}\tau |\frac{\textrm{d}r(\tau)}{\textrm{d}\tau}|^2$, just as in Brownian motion.
However, there are two features that make $r(\tau)$ not literal Brownian motion: $r(\tau)$ is required to be periodic, and the fact that $G_0 = 0$ means that $\int \textrm{d}\tau r(\tau)$ is fixed to be zero.

Evaluating $p_{G0}(h)$ exactly for such a process is quite difficult, but we can approximate it at small $\beta$.
Note that when $G_{\omega} = 0$, $p_{00}(h)$ is simply a step function --- 0 for $h < 0$ and 1 for $h > 0$.
At small but non-zero $G_{\omega}$, such as for small $\beta$, $p_{G0}(h)$ is still (identically) zero for $h < 0$ and still approaches 1 as $h$ increases.
We will approximate it as a step function that transitions at $h = \delta R$, where the latter can be computed from $p_{G0}(h)$ by $\delta R = \int_0^{\infty} \textrm{d}h (1 - p_{G0}(h))$.
One can show that this gives the same answer to leading order when convolved against a slower-varying function, such as we then do to calculate $p_{GA}(h)$.

Evaluating $\int_0^{\infty} \textrm{d}h (1 - p_{G0}(h))$ is in fact much more feasible.
Integrate by parts to obtain $\int_0^{\infty} \textrm{d}h h p_{G0}'(h)$, and since $p_{G0}'(h)$ is the distribution for $\max_{\tau} r(\tau)$ as noted above, this is simply the expected value of $\max_{\tau} r(\tau) \equiv r_{\textrm{max}}$.
The condition that $\int \textrm{d}\tau r(\tau) = 0$ is inconvenient, but we can replace it with the condition that $r(0) = 0$ for the purpose of calculating $\langle r_{\textrm{max}} \rangle_{r,G0}$: first note that $\langle r_{\textrm{max}} \rangle_{r,G0} = \langle (r_{\textrm{max}} - r(0)) \rangle_{r,G0}$, and then by translation invariance we can shift the trajectories being averaged over to each have $r(0) = 0$ (a more formal way of stating this is that each trajectory with $\int \textrm{d}\tau r(\tau) = 0$ is in one-to-one correspondence with a trajectory having $r(0) = 0$ and the same probability under the Brownian measure).
Thus we are left only with evaluating the average of $r_{\textrm{max}}$ under periodic Brownian motion starting and ending at $r(0) = 0$ (sometimes called a ``Brownian bridge'').

\begin{figure}
\centering
\includegraphics[width=0.7\textwidth]{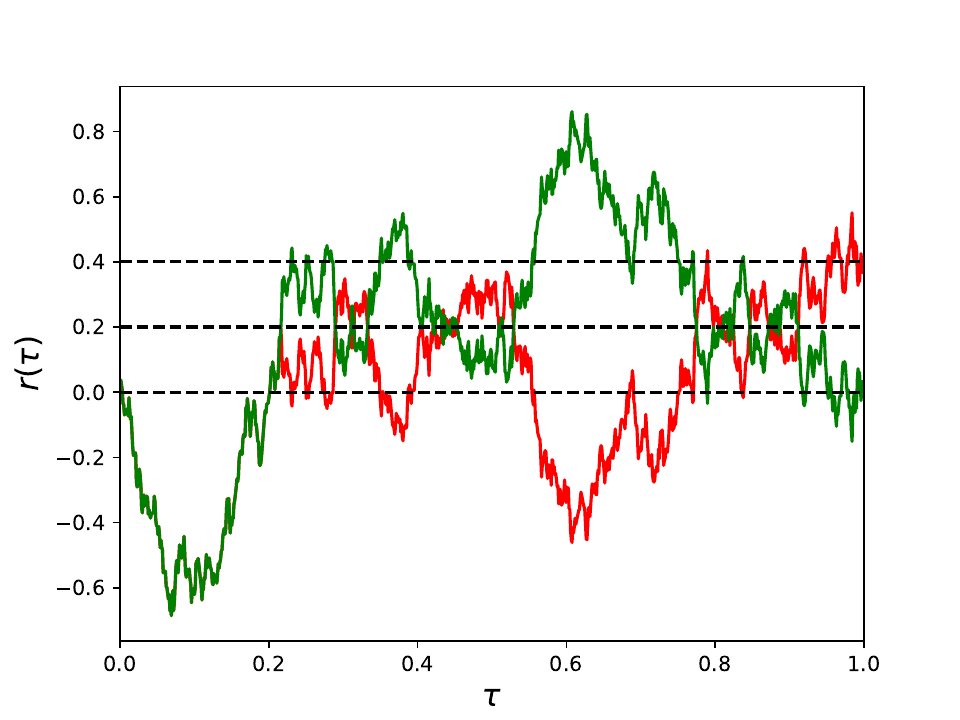}
\caption{An example of a trajectory drawn from a period-1 Brownian bridge (green) and its reflected counterpart used to calculate the average of $r_{\textrm{max}}$ (red). At the first time where the green trajectory crosses $r = 0.2$ (middle horizontal line), we reflect about $r = 0.2$ --- while the green trajectory ends at $r = 0$ (lower line), the red trajectory ends at $r = 0.4$ (upper line).}
\label{fig:images}
\end{figure}

We can compute the distribution of $r_{\textrm{max}}$ under a Brownian bridge using the method of images.
The bridge is equivalently Brownian motion conditioned on the event $r(0) = r(\beta)$.
For any path in this class that first crosses $h$ at some time $\tau_0$ (thus going from 0 to $h$ to 0 as time increases from 0 to $\tau_0$ to $\beta$), there is an equally likely path that is mirrored after $\tau_0$, ending at $2h$ (see Fig.~\ref{fig:images}).
Similarly, any path in Brownian motion that ends at $2h$ can be mapped onto a periodic path by reflecting at the first time where the path crosses $h$.
Thus the probability of a trajectory crossing $h$ and returning to zero equals the probability of a trajectory ending at $2h$, and therefore the cumulative distribution function for $r_{\textrm{max}}$ (under the Brownian bridge) is $1 - \exp [-\Md h^2 / \beta]$.
The expected value of $r_{\textrm{max}}$ is then
\begin{equation} \label{eq:Brownian_bridge_average_max}
\delta R = \int_0^{\infty} \textrm{d}h \frac{2 \Md h^2}{\beta} e^{-\frac{\Md h^2}{\beta}} = \sqrt{\frac{\pi \beta}{4 \Md}}.
\end{equation}

As discussed above, we take $p_{G0}(h)$ to be a step function at $\delta R$.
Thus
\begin{equation} \label{eq:high_temperature_probability_evaluation}
p_{GA}(h) \sim \int_{-h + \delta R}^{\infty} \frac{\textrm{d}z}{\sqrt{2\pi A}} e^{-\frac{z^2}{2A}} = p_{0A}(h - \delta R).
\end{equation}
As a result, the saddle-point equation for $A$ amounts to
\begin{equation} \label{eq:A_saddle_point_equation_high_temperature}
\frac{1}{A} = \frac{\phid}{2} \int \textrm{d}h e^{h - \frac{A}{2}} \frac{p_{0A}'(h - \delta R)^2}{p_{0A}(h - \delta R)} = \frac{\phid e^{\delta R}}{2} \int \textrm{d}h e^{h - \frac{A}{2}} \frac{p_{0A}'(h)^2}{p_{0A}(h)}.
\end{equation}
We see that, compared to the classical case, the effect of quantum fluctuations is simply to increase the filling fraction by an amount $e^{\delta R} = 1 + O(\beta^{1/2})$.
Thus the glass phase occurs at a slightly smaller density, namely 
\begin{equation}
    \phid_c(\beta) \sim \left( 1 - \sqrt{\frac{\pi \beta}{4\Md}} \right) \phid_c(0).
    \label{eq:phiHot}
\end{equation}
It is interesting to note that this correction, although small, is non-analytic in $\beta$ and in particular is larger than what the first analytic correction would be.

\section{Numerical Results} \label{sec:numerics}

The most demanding aspect of determining the phase diagram is solving Eq.~\eqref{eq:G_saddle_point_equations} for $G_{\omega}$, but even this can be done numerically as follows.
Define the ``self-energy''
\begin{equation} \label{eq:self_energy_equation}
\Sigma_{\omega} \equiv -\phid \frac{\partial}{\partial G_{\omega}} \int \textrm{d}h e^{h - \frac{G(0)}{2}} \Big[ p_{G0}(h) - 1 \Big],
\end{equation}
which corresponds to (twice) the last three terms of Eq.~\eqref{eq:G_saddle_point_equations}.
The ``solution'' for $G_{\omega}$ is then simply
\begin{equation} \label{eq:G_formal_expression}
G_{\omega} = \left[ \frac{\Md \omega^2}{2} + \Sigma_{\omega} \right]^{-1}.
\end{equation}
We solve the saddle-point equations iteratively: use an initial guess for $G_{\omega}$ to evaluate $\Sigma_{\omega}$ via Eq.~\eqref{eq:self_energy_equation}, then evaluate an updated $G_{\omega}$ via Eq.~\eqref{eq:G_formal_expression}, which gives an updated $\Sigma_{\omega}$ via Eq.~\eqref{eq:self_energy_equation}, and so on.
While there is no guarantee that such a straightforward iteration must converge, we find that in practice it works quite well, especially when the result for $G_{\omega}$ at slightly higher temperature and/or lower filling fraction is used as an initial guess.

\begin{figure}
\centering
\includegraphics[width=0.7\textwidth]{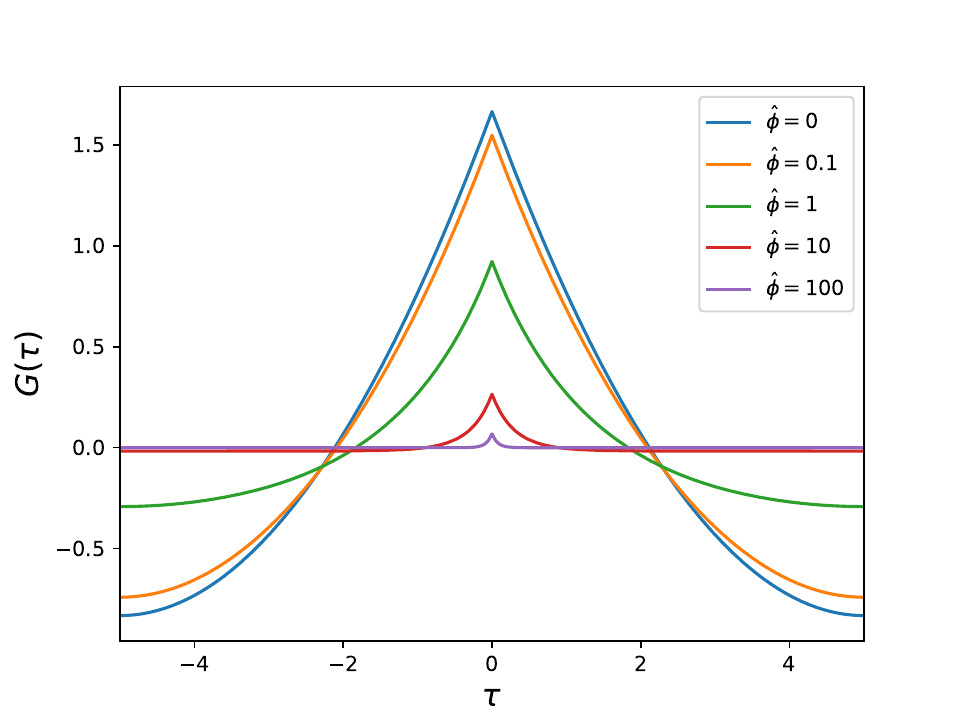}
\caption{$G(\t)$ for various values of $\phid$ at $\beta / \Md = 10$.}
\label{fig:Gt}
\end{figure}

Fig.~\ref{fig:Gt} plots $G(\tau)$ at a rather low temperature for various values of $\phid$ (still specializing to hard spheres).
One can confirm that at $\phid = 0$, the Fourier transform of $G_{\omega}$ is quadratic in $\tau$.
We see in Fig.~\ref{fig:Gt} that $G(\tau)$ begins to better resemble an exponential instead as the filling fraction increases.

\begin{figure}
\centering
\includegraphics[scale=0.5]{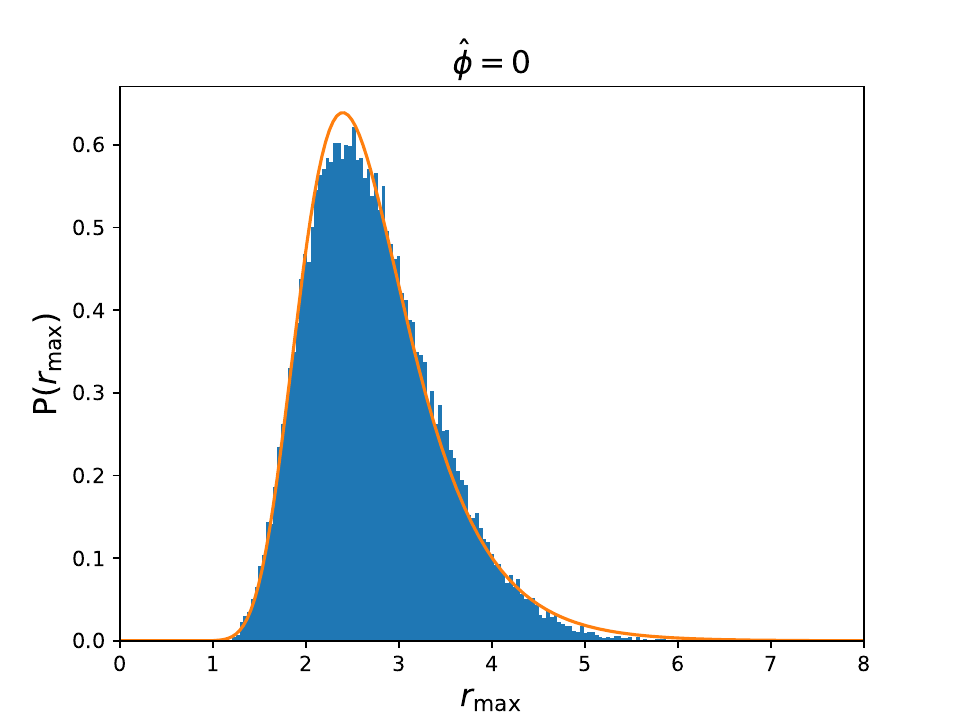}
\includegraphics[scale=0.5]{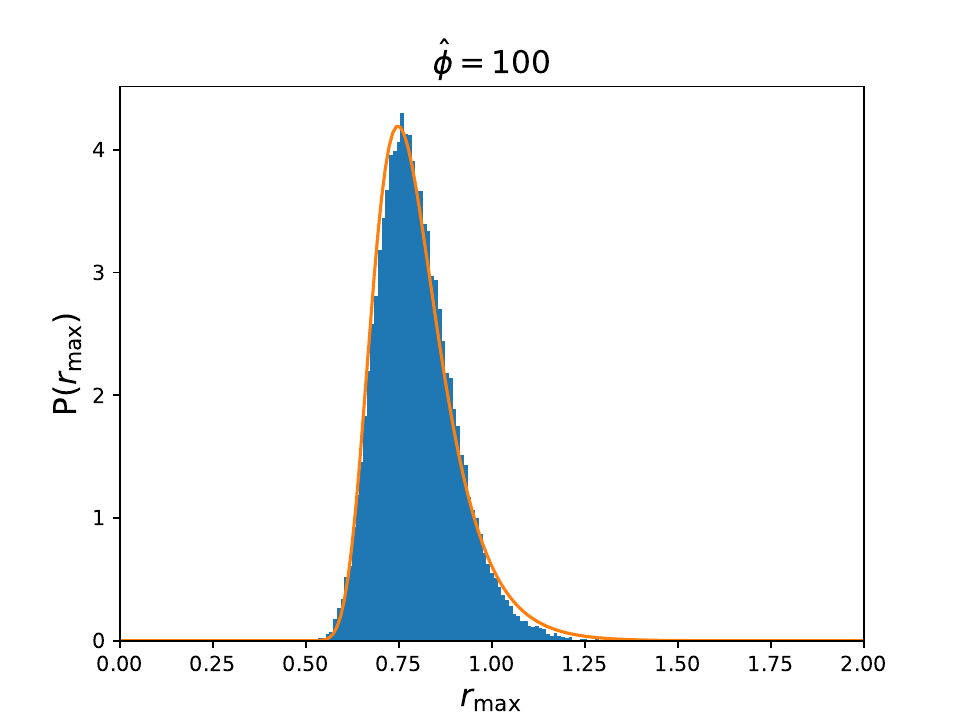}
\caption{Distribution of $r_{\textrm{max}}$ for two choices of $\phid$, representative of small and large filling fractions, both at $\beta / \Md = 10$. Orange curves are best-fit Gumbel distributions. (Left) Small $\phid$, namely the non-interacting limit. (Right) Large $\phid$.}
\label{fig:extreme_value_plots}
\end{figure}

Of particular importance for the saddle-point equations is the extreme-value statistics of a Gaussian process with covariance $G(\tau_{12})$.
When $G(\tau)$ decays exponentially over an interval $\gamma^{-1}$ which is much smaller than $\beta$, this is essentially the extreme-value statistics of $\gamma \beta$ independent Gaussians with variance $G(0)$, which is known to follow the Gumbel distribution: $\textrm{Pr}[r_{\textrm{max}} < x] = \exp [-e^{-(x - \mu)/\sigma}]$ for certain $\mu$ and $\sigma$.
Fig.~\ref{fig:extreme_value_plots} shows the distribution of $r_{\textrm{max}}$ for $\phid = 0$ and $\phid = 100$ alongside best-fit Gumbel distributions.
Although there are systematic deviations, we see that the Gumbel distribution gives a reasonably good fit in both cases (even for $\phid = 0$, where $G(\tau)$ is far from an exponential).

Most importantly, once $G(\tau)$ is computed for a given choice of $\beta$ and $\phid$, we can determine whether any finite value of $A$ solves Eq.~\eqref{eq:A_saddle_point_equation_compact}.
If so, then the system is in the glass phase.
If not, then the system is in the liquid phase.
The resulting phase diagram of quantum hard spheres is given in Fig.~\ref{fig:phaseDiagram}.
First recall that the numerical results agree quite well with those of the high-temperature expansion at small $\beta$ (see Fig.~\ref{fig:smallBetaGraph}).
More broadly, we find that the phase boundary $\phid_c(\beta)$ continues to decrease monotonically at larger $\beta$.
This is in contrast with the results of MCT~\cite{resurgence}, but we again stress that these conclusions are not inconsistent, since MCT relies on a different set of approximations than the high-dimensional limit considered here.

\section{Conclusion} \label{sec:conclusion}

To summarize, in this work we have calculated the phase boundary between the liquid and glass phases of quantum hard spheres, specifically in the infinite-dimensional limit, where a formally exact calculation of the Franz-Parisi potential is possible.
We have determined the leading-order correction to the critical filling fraction $\phid_c$ analytically at small inverse temperature $\beta$, finding that quantum fluctuations give the spheres an effective radius that is enhanced by a multiple of the thermal de Broglie wavelength (and thus $\phid_c$ decreases as $\beta$ increases).
We have also computed the phase boundary numerically at lower temperatures, confirming that $\phid_c$ continues to decrease monotonically.

While it is undeniably valuable to have controlled calculations of the phase diagram in certain limits, such as we have given here, many open questions and directions for future research remain.
First and foremost is investigating the extent to which our results are relevant for actual three-dimensional quantum glasses.
As we have noted, the phase diagram of infinite-dimensional quantum hard spheres stands in some contrast to the predictions of MCT~\cite{resurgence}.
Neither calculation can claim to give a fully accurate description of physical glasses, however --- for one thing, the lifetimes of metastable states are finite outside of mean-field theory and so the liquid-glass phase boundary is (at least in principle) ill-defined.
It would be of great interest to understand which vestiges of the phase diagram remain in realistic systems, and in particular whether the critical filling fraction is indeed monotonic.

Another open question is how the phase diagram is modified for soft-sphere interaction potentials (i.e., potentials that remain finite for separations less than $\ell$).
While we expect it to be qualitatively similar to a certain degree, one new ingredient that soft-sphere potentials add is quantum tunneling through the energy barriers repelling spheres.
One might imagine that this effect suppresses glassiness by allowing the system to better escape from metastable states, and thus there may be an interesting competition between tunneling and the increased effective radius due to quantum fluctuations.

Lastly, it will be important to understand how the tendency to form a glass competes with other types of order at low temperature --- this question is relevant for classical systems as well, but it is especially important here since our calculation suggests that any (hard-sphere) liquid will become a glass at sufficiently low temperature.
Many substances in nature of course do not form glasses, instead entering phases such as solids or superfluids.
Accounting for such effects is beyond the scope of the calculation presented here, but if a suitable generalization can be developed, then it would be quite interesting to study the interplay between these different types of order.

\subsection*{Acknowledgements}

We would like to thank L. Foini, D. Reichman, P. Urbani, and F. Zamponi for valuable discussions and encouragement. This material is based upon work supported by the JQI fellowship (M.W.) and the National Science Foundation Graduate Research Fellowship Program under Grant No. DGE 2236417 (R.B.).  V.G. was supported by the U.S. Department of Energy, Office of Science, Basic Energy Sciences under Award No. DE-SC0001911. We gratefully acknowledge support from Joint Quantum Institute and from AFOSR under grant number FA9550-19-1-0360. The numerical data were obtained using the Zaratan high-performance computing cluster at the University of Maryland, College Park. 

\appendix

\section{Life at large $d$} \label{app:large_d}

The volume of a radius-$\ell/2$ sphere in $d$ dimensions is given by
\begin{equation} \label{eq:single_sphere_volume}
V_d = \frac{\pi^{\frac{d}{2}}}{\Gamma \big( \frac{d}{2} + 1 \big)} \left( \frac{\ell}{2} \right)^d \sim \frac{1}{\sqrt{\pi d}} \left( \frac{\pi e \ell^2}{2d} \right)^{\frac{d}{2}},
\end{equation}
where $\Gamma$ denotes the gamma function.
Note that the volume shrinks even faster than exponentially, in particular faster than $\ell^d$.
This means that in large dimensions, the most obvious packing of spheres --- a simple cubic lattice --- only fills an infinitesimal fraction of space.

More efficient packings exist with far larger filling fractions --- this is in fact a well-studied problem~\cite{rogers1964packing,1947,venkatesh}.
Hermann Minkowski proved the first real lower bound on the maximal filling fraction, that in any dimension a fraction of $2^{-d}$ is possible.
The proof is elementary: given a packing, if we double the radius of each sphere ($r \rightarrow 2r$), then either every point in space becomes covered by an expanded sphere or there is a point whose distance from the center of every sphere is at least $2r$.
In the former case, the original filling fraction must be at least $2^{-d}$.
In the latter case, the packing must not have been optimal (because a sphere of radius $r$ can be inserted at the aforementioned point).

More advanced proofs have established that packings exist with filling fractions $\phi$ at various multiples of $d 2^{-d}$.
Thus it is natural to write $\phi = d2^{-d} \phid$ and consider $\phid$ finite at large $d$.
It is interesting to note that the analysis in this paper can apply to arbitrarily large $\phid$, even though packings at sufficiently large $\phid$ haven't been rigorously proven to exist (although they have been established for $\phid$ well beyond where we locate the liquid-glass phase boundary~\cite{venkatesh}).

The search for upper bounds on allowed filling fractions has been even more difficult.
The best known bounds scale as $2^{-cd}$ for a constant $c \approx 0.599$, with hopes to possibly improve to $c \approx 0.779$, but anything further would require substantial breakthroughs~\cite{Cohn_2014}.

Another consequence of the scaling of $V_d$ in Eq.~\eqref{eq:single_sphere_volume} is that we can ignore the effects of particle exchange and in particular the possibility of Bose-Einstein condensation.
The densities that we consider in this paper are of order $d / 2^d V_d$, which grows predominantly as $d^{d/2}$.
On the other hand, the density scale for condensation is of order $\lambda_T^{-d} \sim (M/\beta)^{d/2}$, where $\lambda_T$ denotes the thermal de Broglie wavelength.
Since we take $M$ to be of order $d^2$, the condensation density grows as $d^d$, much larger than the critical density for glass formation.

It is also worth asking how many spheres lie within a distance $\ell (1 + \delta)$ from the center of a given sphere (note that two spheres are in contact when they are separated by distance $\ell$).
Multiplying the density by the volume of a radius-$\ell$ sphere (not $\ell/2$), we find that the number scales as $d e^{d \delta}$.
Thus the number of spheres whose edges are within $O(1/d)$ is $O(d)$, i.e., large.
This motivates the interpretation of the infinite-dimensional liquid as a mean-field system --- in general, systems in which each degree of freedom interacts with a large number of neighbors are well-described by mean-field theory.

\bibliographystyle{ieeetr}
\bibliography{main.bib}
\end{document}